# Lithium Intercalation in Graphene/MoS$_2$ Composites: First-Principles Insights

Xiji Shao, Kedong Wang, Rui Pang, and Xingqiang Shi*

*Department of physics, South University of science and technology of China, Shenzhen 518055, China*

[*]E-mail: shixq@sustc.edu.cn

**Abstract**

As a storage material for Li-ion batteries, graphene/molybdenum disulfide (Gr/MoS$_2$) composites have been intensively studied in experiments. But the relevant theoretical works from first-principles are lacking. In the current work, van-der-Waals-corrected density functional theory calculations are performed to investigate the interaction of Li in Gr/MoS$_2$ composites. Three interesting features are revealed for the intercalated Gr/Li(n)/MoS$_2$ composites (n = 1 to 9). One is the reason for large Li storage capacity of Gr/MoS$_2$: due to the binding energies per Li atom increase with the increasing number of intercalated Li atoms. Secondly, the band gap opening of Gr is found, and the band gap is enlarged with the increasing number of intercalated Li atoms, up to 160 meV with nine Li; hence these results suggest an efficient way to tune the band gap of graphene. Thirdly, the Dirac cone of Gr always preserve for different number of ionic bonded Li atoms.

**TOC Graphic:**

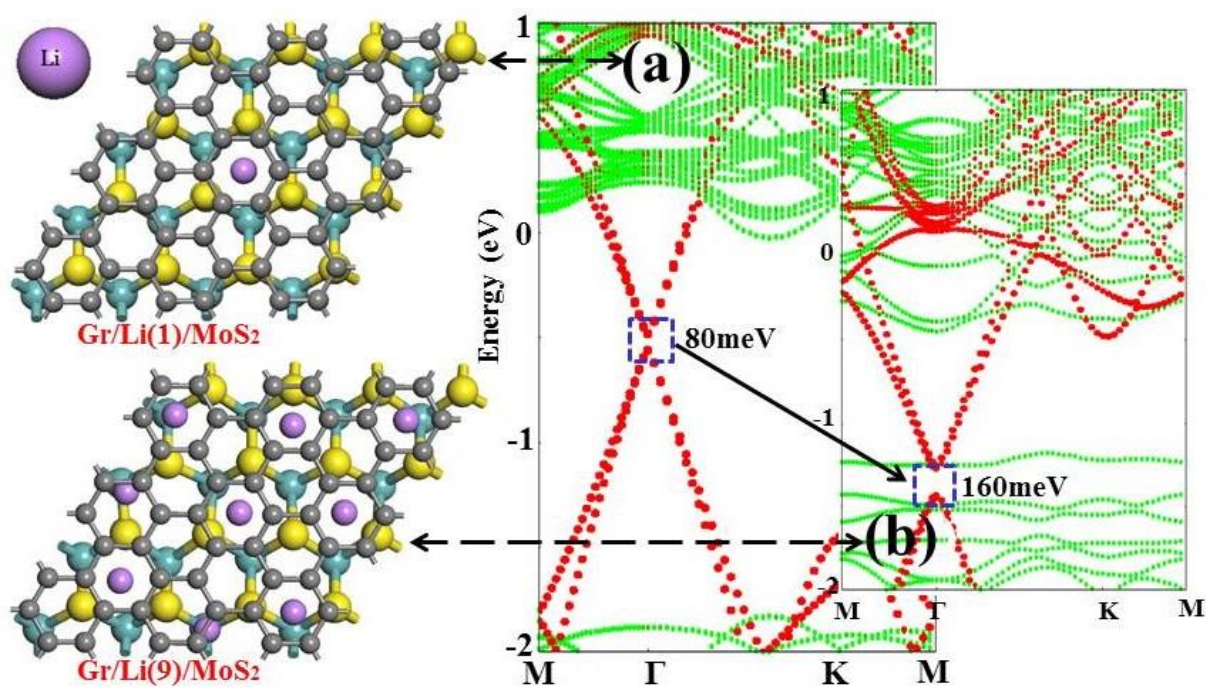

**Keywords**: Graphene and MoS$_2$, first-principles calculations, band gap tuning, Dirac cone, Li-ion storage

# 1. Introduction

Graphene (Gr) has been intensively studied due to its extraordinary properties, such as ultra-high charger mobility and electrical conductivity, and large surface area etc. [1-4]. But it is a zero-gap semiconductor [5-7], which is a disadvantage for electronic and optoelectronic devices applications. Band gap opening of Gr can be achieved by doping, strain, or controlling the Gr dimensionality, etc. [8-10]. However, these methods usually damage the crystallographic structure of Gr and unexpected loss in the excellent electrical properties [11-13]. The Dirac cone of Gr could be preserved under weak van-der-Waals interactions or ionic bonds, while the Dirac cone would be disrupted by strong band hybridization as a result of covalent bonding[14]. Bulk molybdenum disulfide ($MoS_2$) is a semiconductor material with an indirect band gap. It changes to direct band gap when the thickness of the $MoS_2$ slab is reduced to form monolayer [15]. Monolayer $MoS_2$ is composed of three atomic layers stacked with a sequence of S-Mo-S, with Mo and S atoms covalently bonded [16]. For the Gr-($4\times4$)/$MoS_2$-($3\times3$) composites, small band gaps of Gr, up to 30 meV, are opened at the $K$-point of the Brillouin zone [15].

In the aspect of Li-ion batteries, the storage materials should have the following properties -- low cost, large surface areas for Li storage and large charge transfer to Li. Gr/$MoS_2$ composites have unique electronic and structural properties [15], which can provide large space for Li atoms intercalation, and Li intercalation increases the electron conductivity and electrochemical performance of Gr/$MoS_2$ [17]. The Li-ion batteries depend mostly on the electrochemical properties of anode materials. The Gr/$MoS_2$ composites enhanced the electrochemical properties for Li storage [18] and had the excellent Li storage capacities [19-23]. The capacities of the Gr/$MoS_2$ composites were higher than the other components and had excellently cycle stabilities (> 100 cycles), which suggested that they could be used as anode materials for Li-ion batteries [17-18, 24]. The two dimensional (2D) Gr/$MoS_2$ composite have high reversible capacity of 1060 mAh $g^{-1}$ and excellent cycle stability [25]. Chang *et al*. synthesized Gr/$MoS_2$ composites that exhibited excellent capability, up to 1300 mAh$g^{-1}$ and cycling stability for Li-ion batteries [17-19, 24]. Intercalation of atoms or molecules into layered materials shows different physical and chemical properties in comparison with those of pristine materials. Li-intercalated in graphite and few layer graphene, and their electronic structures have been widely studied both experimentally [26-29] and theoretically [30-33].

However, up to now, the physical and chemical properties of Li atoms intercalated in the Gr/$MoS_2$

composites have been seldom reported at the atomic detail from first-principles studies [34]. There is also a lack of theoretical study on different Li concentrations in Gr/MoS$_2$. In the current work, the properties of different number of Li atoms intercalation in Gr/MoS$_2$, including geometric structures, binding energies, charge transfer, band structures and density of states (DOS) were investigated by van der Waals (vdW) corrected density functional theory (DFT) calculations.

2. **Theoretical methods**

All calculations are carried out by the plane-wave-basis-set Vienna ab initio simulation package (VASP) [35] with the van-der-Waals-corrected density functional (vdW-DF). The optB88-vdW-DF [36-37] is used, in which an optimized version of the B88 exchange functional are adopted in the vdW-DF [38], which yield accurate equilibrium interatomic distances and energies for a wide range of systems [39]. The plane wave energy cut-off is 400 eV. The Mo semi-core 4$p$ and 5$s$ states are included as valance electrons. Core electrons are described by projector augmented wave (PAW) potentials [40]. The vacuum space is set to 12 Å. Γ-point Brillouin zone sampling is used for structure optimization, and 8 × 8 × 1 Γ-centered $k$-points sampling are used for electronic structure calculations with the Monkhorst-Pack scheme [41]. The optimized lattice constants of graphene and MoS$_2$ monolayer are 2.465Å and 3.185Å, respectively. The supercell structure of the composite system is Gr-(3√3 × 3√3)R30º /MoS$_2$-(4 × 4), which diminishes the lattice mismatch between Gr and MoS$_2$ layers to only about 0.5%. All atoms are fully relaxed until the residue force on each atom is less than 0.02 eV/Å. Bader charge analysis [42] is used to analyze the charge transfer behaviors among Gr, Li, and MoS$_2$. The monolayer MoS$_2$ with a direct band gap character, which is 1.68 eV in our calculation, similar to that calculated by Abbas Ebnonnasir *et al.* [43] and Lu *et al.* [44].

3. **Results and Discussions**

**3.1 Geometric structures and binding energies**

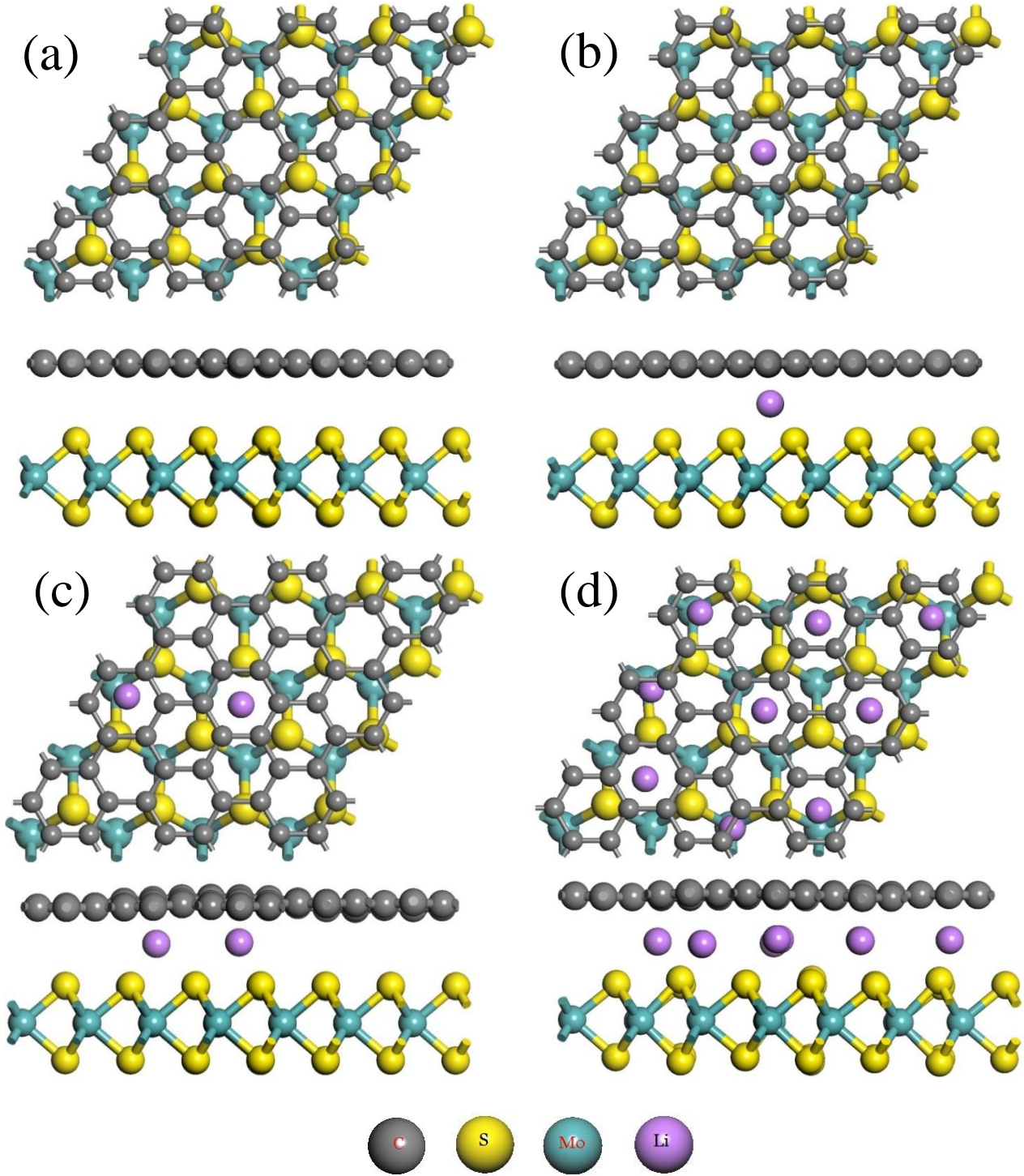

**Fig. 1.** Top and side views of the composite structures: Gr/MoS$_2$ (a), Gr/Li(1)/MoS$_2$ (b), Gr/Li(2)/MoS$_2$ (c), and Gr/Li(9)/MoS$_2$ (d).

First, we study the adsorption behavior of graphene on MoS$_2$. The binding energy is defined as $E_\text{b} = E_\text{Gr} + E_{\text{MoS}_2} - E_{\text{Gr/MoS}_2}$. The optimized structures are depicted in Fig. 1a. The height difference between graphene and the top layer S atoms of MoS$_2$ is 3.37 Å. The binding energy between Gr/MoS$_2$ is 1.52eV per cell (or 0.03 eV per C atom), which is small due to the Gr on MoS$_2$ is in physical adsorption. The results are in agreement with that of Ma *et al.* [45]. The weak interaction between Gr and MoS$_2$ had also been verified by the experiment [46].

Next, we focus on the intercalation behavior of Li in the Gr/MoS$_2$ composite. Different Li concentrations, Gr/Li(n)/MoS$_2$, are considered, where n = 1 to 9 is the number of intercalated Li atoms. The binding energy per Li-atom is defined as

$$E_{\mathrm{b}} = \left(E_{\mathrm{Gr/MoS_2}} + \mathrm{n} \times E_{\mathrm{Li}} - E_{\mathrm{Gr/Li(n)/MoS_2}}\right)/\mathrm{n}, \qquad (1)$$

where $E_{\mathrm{Gr/MoS_2}}$, $E_{\mathrm{Li}}$, and $E_{\mathrm{Gr/Li(n)/MoS_2}}$ are total energies of Gr/MoS$_2$, of a single Li atom in gas phase or in Li-bulk phase, and of the Gr/Li(n)/MoS$_2$ composite systems, respectively.

For one Li atom interaction in Gr/MoS$_2$, we calculate the different situations such as Li atom locates on the surface of Gr (Li/Gr/MoS$_2$) and on the surface of MoS$_2$ (Gr/MoS$_2$/Li). We found that both Li/Gr/MoS$_2$ and Gr/MoS$_2$/Li are less stable than Li intercalated in Gr/MoS$_2$ to form Gr/Li/MoS$_2$. The latter intercalated structure is more stable than the former two surface adsorption structures by about one eV. Then, we focus on the intercalated structure Gr/Li/MoS$_2$ in the following.

For the intercalated structure, different models are tried and we find that Li tends to at the 'hollow' sites of Gr and MoS$_2$, keeping the Gr-MoS$_2$ vertical distance close to about 3.37 Å as that in the pure Gr/MoS$_2$ structure without Li. For MoS$_2$, the 'hollow' sites also include the Mo top sites. So, two models are preferred: (i) the intercalated Li atom at the hollow sites of both MoS$_2$ and Gr and (ii) the Li atom at the top of Mo atom and at the hollow of Gr. The binding energies of these two models are about 1.73 eV and 1.35 eV, respectively. It suggests that the Li atom prefers to form model (i), see Fig. 1b. In the structure shown in Fig. 1b, the average vertical height between the Gr and MoS$_2$ layers remains 3.37 Å, almost identical to the height without Li (Fig. 1a). The height of 3.37 Å with one Li is similar to the result of Ref. [34]. The average bond length is 2.28 Å between Li and the six nearest C atoms, and that is 2.48 Å between Li and the three nearest S atoms. The maximum corrugation in the graphene layer is about 0.09 Å.

With the number of Li atoms increasing from one to two, the Mo on top sites becomes possible; with even more Li atoms, the C–C bridge sites also become possible. However, a large proportion of Li atoms are located at the sites of the above discussed model (i) and model (ii). For the interacted number of Li from two to nine, the distance of the graphene and MoS$_2$ increases from 3.49 to 3.75 Å (see Fig. 2a), i.e. more Li atoms cause a larger Gr–MoS$_2$ layer separation. For two Li atoms intercalation in Gr/MoS$_2$, the average bond length of Li-C and Li-S are 2.33 Å and 2.48 Å, respectively. For nine Li atoms, the average bond length of the Li-C and Li-S are 2.44 Å and 2.54 Å. Hence the distance of the graphene layer is farther separated from the MoS$_2$ layer with increasing number of intercalated Li.

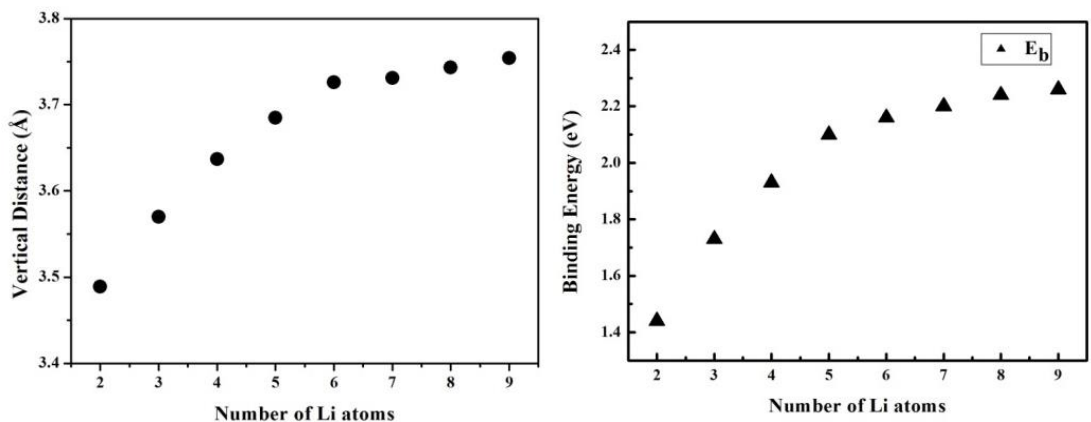

**Fig. 2.** (a) The average vertical distance between graphene and MoS$_2$ as a function of number of intercalated Li atoms; (b) the binding energy per Li as a function of number of intercalated Li atoms.

The binding energies per Li for different Li concentrations are shown in Fig. 2b, which used the Li atom in gas phase as a reference, see the previous equation (1) defining the binding energy per Li-atom. For Li-bulk as a reference, nothing changes in the binding energy curve except a rigid shift of about 0.7 eV. The binding energies per Li increase with the number of Li increasing, and the increase is smaller at the nine-Li end (Fig. 2b). The reason is as following. With more Li atom intercalation, the preferred Gr-MoS$_2$ vertical separation with Li is about 3.75 Å (for 9 Li), while with few Li atoms the Gr-MoS$_2$ vertical separation is decreased significantly (see Fig 2a, for less than six Li), which means that Li atoms are compressed by Gr/MoS$_2$ in the few Li atom cases. The compression to Li makes the Li binding energy decrease. And the rate of change in the binding energy is proportional to the rate of change in the vertical separation (compare Fig. 2b to 2a). From this trend, it can be inferred that with more Li atoms intercalation, the binding energies will not change much. In another words, the binding energy will achieve a maximum value of about 2.3 eV per atom, with tiny fluctuations for the number of Li atoms beyond nine. Then, we conclude that the binding energy per Li will not depend on the Li concentration beyond the nine Li case. The graphene layer has the maximum distortion from 0.45 Å (for two Li) to 0.2 Å (for nine Li), it is flatter at the nine Li case in comparison to the fewer Li cases; this is another reason for the binding energy increase.

**3.2 Electronic structures**

We quantify the charge transfer between graphene, Li, and $MoS_2$ through Bader charge analysis. The results are listed in Table 1. In this table, a minus value means lose electrons. The electron transfer occurs when graphene and $MoS_2$ are brought together. Due to physical adsorption of Gr on $MoS_2$, the charge transfer is small -- the Gr layer lose only 0.05 electrons per unitcell. However, in the cases with Li intercalations, electron transfer from Li to both Gr and $MoS_2$ is siginificant. The reason is that ionic bonds being formed between Li-C and Li-S. The Gr layer gain additional 0.38 to 3.36 electrons from the one Li to the nine Li cases in comparison to the case without Li, while the S-top layer gain additional 0.47 to 4.64 electrons from the one Li to nine Li cases. The larger the number of intercalated Li, the more the transferred charge. Meanwhile, the S-top layer of $MoS_2$ gain more electrons than the Gr layer by about 0.09 to 1.28 electrons for the number of Li from one to nine. So there's a net dipole point from the S-top layer to the Gr layer with Li intercalation (in comparison to the case without Li), see Schematic 1; and this dipole induces a dipole with a opposite direction, point from the Mo layer to the S-down layer; this explains why the S-down atoms lose electrons with Li intercaations compared to that without Li, as shown in Table 1.

**Table 1.** Number of transferred electrons per Gr-(3√3×3√3)/Li(n)/$MoS_2$-(4×4) cell among the Gr, Li, and $MoS_2$ layers, as a function of number of intercalated Li atoms from n = 0 to 9. A minus value means lose of electrons. S-top is the S layer close to Li.

|        | 0      | 1      | 2      | 3      | 4      | 5      | 6      | 7      | 8      | 9      |
|--------|--------|--------|--------|--------|--------|--------|--------|--------|--------|--------|
| Gr     | -0.05  | 0.33   | 0.66   | 0.99   | 1.49   | 1.89   | 2.28   | 2.62   | 2.98   | 3.31   |
| Li     | 0      | -0.87  | -1.75  | -2.62  | -3.49  | -4.36  | -5.25  | -6.13  | -7.00  | -7.87  |
| $MoS_2$ | 0.05  | 0.53   | 1.09   | 1.63   | 2.01   | 2.47   | 2.97   | 3.51   | 4.02   | 4.56   |
| S-top  | 8.74   | 9.21   | 10.03  | 10.76  | 11.03  | 11.31  | 11.94  | 12.39  | 12.84  | 13.38  |
| Mo     | -18.24 | -17.88 | -17.94 | -18.28 | -18.02 | -17.87 | -18.06 | -18.04 | -17.98 | -17.91 |
| S-down | 9.55   | 9.20   | 9.00   | 9.15   | 9.00   | 9.03   | 9.09   | 9.15   | 9.16   | 9.07   |

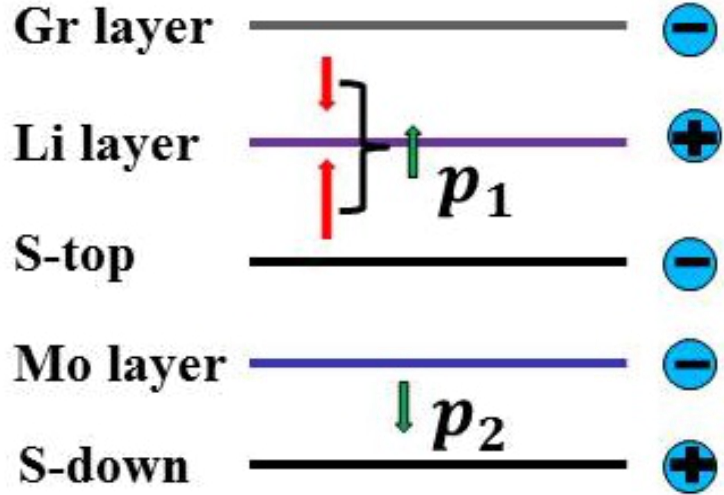

Schematic 1: The net dipole $p_1$ point from the S-top layer to the Gr layer with Li intercalation (in comparison to the case without Li, see Table 1), which induced a inversed dipole $p_2$ point from the Mo layer to the S-down layer.

Pure graphene retains its semimetal character and the Dirac cone emerges at the $\Gamma'$ point (Fig. 3a) for the $(\sqrt{3} \times \sqrt{3})R30°$ surpercell due to the Brillouin zones folding, which is different to the $1 \times 1$ cell, to which the Dirac cone emerges at the $K$ point[43]. For Gr in the hybrid system Gr/MoS2, the opened gap is about 0.008 eV between $\pi$ and $\pi^*$ states (Fig. 3b). The gap opening can be attributed to the MoS$_2$ interaction, which leads to the symmetry breaking of Gr [14, 47-48]. Fig. 3c and 3d show that the Dirac points shift down to below the Fermi level by about 0.40 eV and 1.28 eV for the one Li and nine Li cases, respectively. And the energy gap of Gr is enlarged to about 0.08 eV for the one Li case and 0.16 eV for the nine Li case, respectively. The Li atoms intercalation induces local corrugations in Gr, which leads to an alteration from local $sp^2$ to partial $sp^3$ bonding in Gr; this further destroys the symmetry of Gr and generates a larger band gap. Also, the build-in electric field vertical to the Gr plane is enhanced in the nine Li case due to charge transfer (See Table 1). These results suggest that the ionic bonds between Gr and Li, and the accompany charge transfer are important to open a larger band gap. The contributions of Gr to the energy bands of the full Gr/Li(n)/MoS$_2$ systems are shown with red circles in Fig. 3b-3d, where the size of circles are proportional to their contributions. The Dirac point is always preserved in all cases. The bands near the Dirac points are totally contributed from Gr in all cases. Both the Gr and the MoS$_2$ energy bands cross the Fermi level in the band structure of Gr/Li(n)/MoS$_2$ for n > 1. This characteristic suggests that electrons can be conduct in both the graphene and the MoS$_2$ layers.

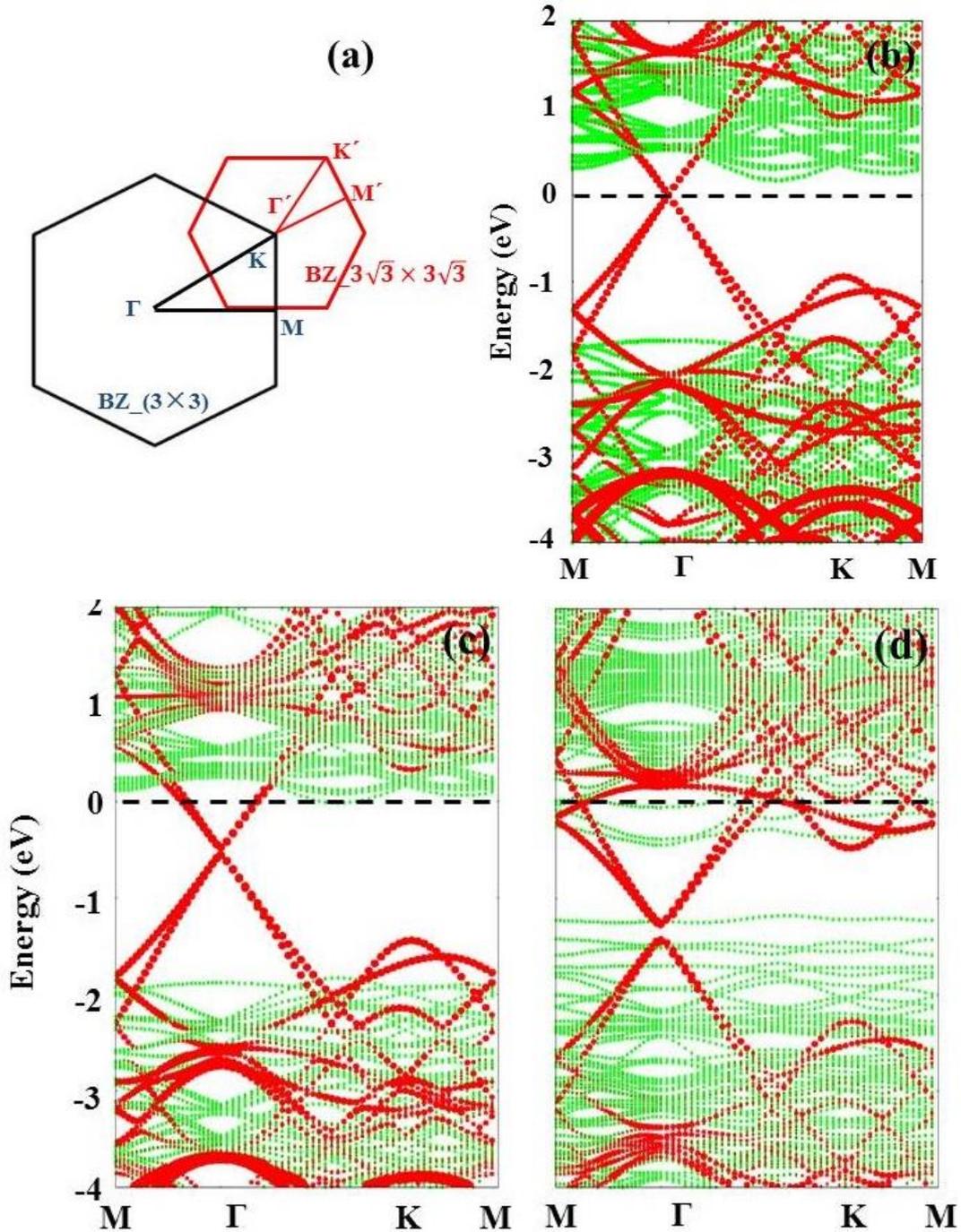

**Fig. 3.** (a) Brillouin zones of (3×3) (larger hexagon in black) and (3√3×3√3)R30° (smaller hexagon in red) graphene unit cells, in which the *K* point of the former is folded to the Γ point of the latter. The band structure of Gr/MoS$_2$ (b), Gr/Li(1)/MoS$_2$ (c), and Gr/Li(9)/MoS$_2$ (d); contributions from graphene to the band structure are shown in red solid circles.

The energy dispersion of the clean graphene is linear near the Dirac cone, and can observe the massless fermion. However, for the Gr/Li(n)/MoS$_2$ systems, the charge redistribution breaks the equivalence sub-lattice in the graphene layer, giving rise to the non-zero bandgap, and the energy dispersion relation around the Dirac cone is not strictly linear, suggesting a massive fermion. The effective mass of carriers at the Dirac point are calculated by the expression $m^* = \frac{\hbar^2 dk^2}{d^2 E}$ [49]. As shown in Table 2, massless or almost massless carriers exit in the Gr/MoS$_2$ and Gr/Li(n)/MoS$_2$ systems.

**Table 2** Effective masses at the Dirac point of the systems ($m_h^*$ represents the bands at the higher energy side of the band gap, while $m_e^*$ represents the bands at the lower energy side).

| System | $m_h^*(m_e)$ | $m_e^*(m_e)$ |
|---|---|---|
| Gr/MoS$_2$ | Γ → M(0.0001) | Γ → M(0.0001) |
|  | Γ → K(0.0001) | Γ → K(0.0001) |
| Gr/Li(1)/MoS$_2$ | Γ → M(0.0003) | Γ → M(0.0003) |
|  | Γ → K(0.0003) | Γ → K(0.0002) |
| Gr/Li(9)/MoS$_2$ | Γ → M(0.0018) | Γ → M(0.0017) |
|  | Γ → K(0.0016) | Γ → K(0.0011) |

To better understand the orbit contribution from Gr, Li, and MoS$_2$, we plot the total density of states (TDOS) and local density of states (LDOS) in Fig. 4. Due to Gr ionically bonded to Li atoms in the composite systems, the Dirac point feature in the LDOS of Gr was not disrupted. For the Gr/MoS$_2$ composite (the black line shown in Fig. 4b), the Dirac point of graphene is slightly shifted to the conduction bands because of a smaller charge transfer as noted in Table 1. The major electron contribution to conduction bands are from *p* (C atoms and S atoms) and *d* orbits (Mo atoms); while the electron contribution from Mo *d* orbits is larger than the C and S *p* orbits.

For Gr/Li(n)/MoS$_2$, n=1 or 9, the total density of states shifts to the valance bands in comparison to Gr/MoS$_2$ without Li, as shown in the total density of states in Fig. 4a. The Dirac point of Gr is shifted to lower energies below the Fermi level, as shown in Fig. 4b. These indicated that electrons transfer to graphene in accordance with Table 1. For C, Mo and S atoms, the mainly electron contributions to valence bands are from *p*, *d* and *p* orbits, respectively, as shown in Fig. 4b-4d. With the number of Li atoms increasing in Gr/MoS$_2$, the Dirac point shifts-down further relative to the Fermi level.

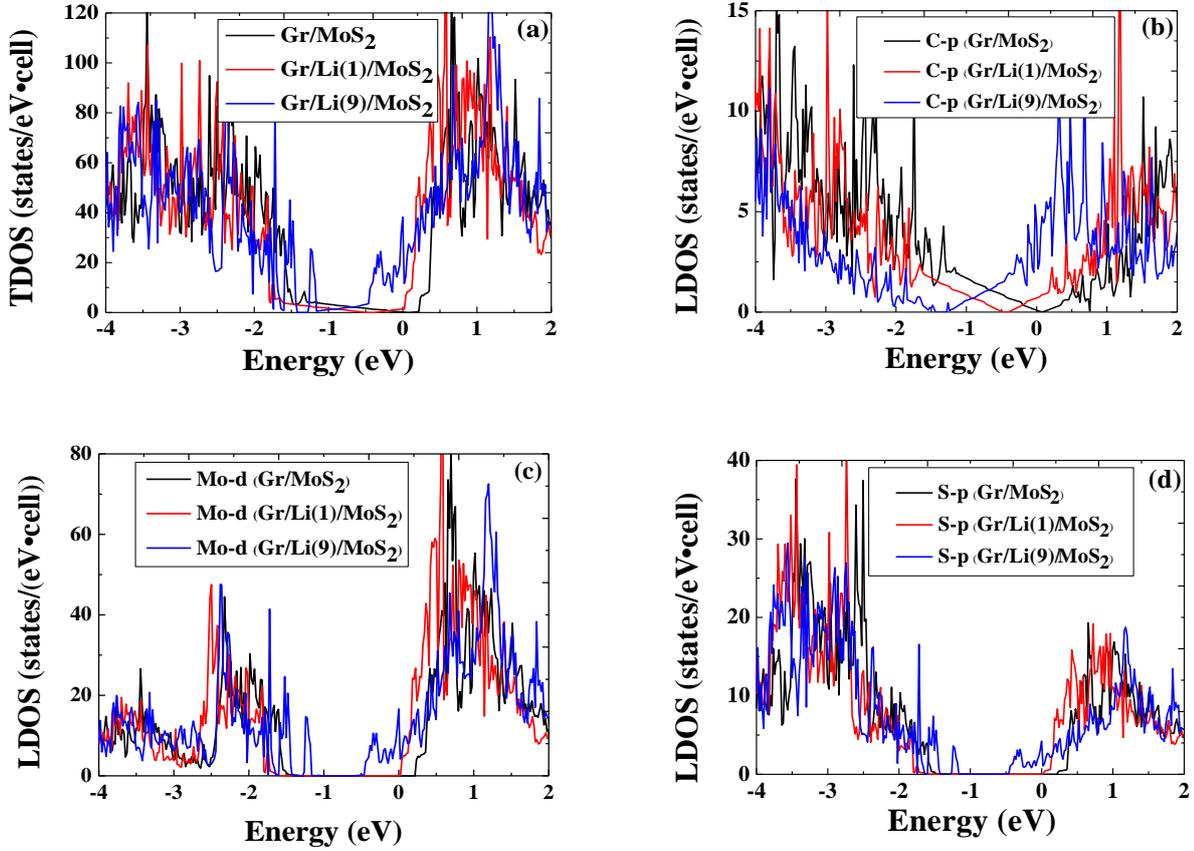

**Fig. 4.** Density of states for the Gr/MoS$_2$, Gr/Li(1)/MoS$_2$ and Gr/Li(9)/MoS$_2$, respectively: the TDOS are shown in (a); The LDOS of C, Mo and S atoms are shown in the (b), (c) and (d), respectively. The Fermi level is set to zero.

## 4. Conclusion

In summary, structural and electronic properties of Gr/Li(n)/MoS$_2$ composites are systematically investigated by vdW-DF calculations. The Gr/MoS$_2$ composites can enhance the Li storage capacity due to the binding energy per Li increase with the increasing number of intercalated Li atoms. The graphene band gap is opened and gap width can be tuned by the number of Li atoms, due to Li atoms intercalation produce the ionic bond, which leads to symmetry breaking of Gr and the charge transfer induced vertical electric field to the Gr plane. The band gap of graphene is enlarged with more intercalated Li atoms, while the Dirac cone feature of graphene is always preserved for Gr/Li(n)/MoS$_2$. Such a bandgap tuning of Gr is essential for its electronics applications.

## Acknowledgments

This work is supported by National Natural Science Foundation of China (Grants 11474145 and 11334003). The authors thank Yangfan Shao, Yan Liu and Qian Wang for helpful discussions. We also thank the National Supercomputing Center in Shenzhen for providing computation time.